\documentstyle[emulateapj,psfig]{article}

\newcommand{\HII}{H{\sc ii}\, \,}

\newcommand{\HeII}{He{\sc ii}\, \,}
\newcommand{\HeIII}{He{\sc iii}\, \,}

\def\msun{\,{\rm M_\odot}}

\newcommand{\etal}{et~al.\ }
\righthead{Very Massive Stars}

\makeatletter

\newenvironment{figurehere}
  {\def\@captype{figure}}
  {}
\makeatother

\begin{document}

\title{Did Very Massive Stars Pre--enrich and Reionize the Universe?}
\lefthead{Oh, Nollett, Madau \& Wasserburg}
\righthead{Did Very Massive Stars Pre--enrich and Reionize the Universe?}

\author{S. Peng Oh\altaffilmark{1},
Kenneth M. Nollett\altaffilmark{1}, Piero Madau\altaffilmark{2},
and G.J. Wasserburg\altaffilmark{3}}
\altaffiltext{1}{Theoretical Astrophysics, Mail Code 130-33, Caltech, Pasadena, CA~~91125}
\altaffiltext{2}{Department of Astronomy and Astrophysics, University of
California Santa Cruz, Santa Cruz, CA 95064}
\altaffiltext{3}{The Lunatic Asylum, Division of Geological and
Planetary Sciences, Caltech, Pasadena, CA~~91125}

\begin{abstract}
\noindent Recent studies of heavy $r$--process elements in low [Fe/H]
halo stars have suggested that an initial population of metal-free
very massive stars (VMSs) may be required to provide early Fe
enrichment without coproducing heavy $r$ nuclei.  We find similar
abundance trends in $\alpha$--elements (which should be copiously
produced by VMSs), but not in other elements such as carbon (which
should not), in agreement with this hypothesis.  We then combine the
corresponding level of prompt initial enrichment with models of VMS
nucleosynthetic yields and spectra to estimate the corresponding ionizing fluxes.  The result suggests that
there may have been enough VMS activity to reionize the universe.  The
unusually hard spectrum of VMSs would imply a different reionization
history from canonical models. \HeII could have been reionized at high
redshift, only to recombine as a subsequent generation of stars formed
with a ``normal'' initial mass function.

\end{abstract}

\section{Introduction}

The nature of the very first generation of metal--free (`Population
III') stars has been debated for decades (Schwarzschild \& Spitzer
1953; Ezer \& Cameron 1971).  It is often thought that the lack of efficient cooling
mechanisms would lead to a top--heavy initial mass function (IMF)
(Larson 1998), and particularly to the production of `very massive
stars' (VMSs) with ${\rm M_{*}} >100\,$M$_\odot$ (Carr, Bond \& Arnett 1984).
Recent numerical simulations of cooling primordial molecular gas have
hinted that this is indeed the case: they find
quasistatically-contracting gas clumps with characteristic densities
$\sim 10^{4} \, {\rm cm^{-3}}$ and temperatures of a few hundred
kelvin -- a regime set by the properties of the chief coolant,
molecular hydrogen (Bromm, Coppi, \& Larson 2001a; Abel, Bryan \&
Norman 2000; Nakamura \& Umemura 2001).  The Jeans unstable clumps are
of order a few hundred solar masses; because the contraction is
subsonic, further fragmentation is unlikely, suggesting that the
very first stars were indeed very massive. Among other properties, VMSs would produce many more ionizing photons
per unit stellar mass than ordinary stars (Bromm, Kudritzki, \& Loeb
2001c), and have much harder spectra; they could thus ionize \HeII
(Tumlinson \& Shull 2000). Their existence would have important
implications for the metal enrichment and reionization history of the
universe.

The duration and extent of early metal--free star formation are not
known, and depend on the complex and inhomogeneous process of early
metal enrichment (Gnedin \& Ostriker 1997; Madau, Ferrara, \& Rees
2001). At present,
the best evidence for VMSs with ${\rm M_{*}}<250\,$M$_\odot$ may be from
their distinctive nucleosynthetic products: VMSs process more of their
mass into alpha-elements than normal stars, and they finally explode
by a mechanism (pair instability) that is very different from core-collapse
supernovae (Fryer \etal 2001), and which is unlikely to produce $r$-process elements (Heger \& Woosley 2001, hereafter HW). Metal-poor
halo stars provide important fossil evidence for early nucleosynthesis
because they extend to metallicities even lower than those inferred
for the damped Ly$\alpha$ systems (e.g., Ryan 2000). 

Recent observations of $r$--process abundances as a function of [Fe/H]
in halo stars suggest that an initial population of rapidly evolving
VMSs provided an intial Fe inventory without co--producing heavy
$r$--nuclei (Wasserburg \& Qian 2000, hereafter WQ; Qian \& Wasserburg 2001b).  In this {\it Letter}, we
estimate the total number of VMSs necessary to produce the required
pre-enrichment, and consider the consequences for
reionization, in the spirit of previous papers (e.g. Haiman \& Loeb 1997, Miralda-Escud\'e \& Rees 1998) which have
considered the connection between the metallicity of the Ly$\alpha$
forest at $z \sim 3$ and high-redshift ionizing photon production for
a 'normal' IMF. In all numerical estimates, we assume a $\Lambda$CDM cosmology with
$(\Omega_M,\Omega_\Lambda,\Omega_b,h, \sigma_{8
h^{-1}},n)=(0.35,0.65,0.04,0.65,0.87,0.96)$.

\section{Enrichment by very massive stars}

In Figure \ref{yields} we show VMS yields as functions of mass, as
computed by HW for stars in the mass range subject to pair instability
explosions $M_{*} \approx$ 140-260 $M_{\odot}$. Such stars are
completely disrupted and leave no remnants. Stars with $40\, {\rm
M_{\odot}}< {\rm M}_{*} < 100\,$M$_\odot$ and $M_{*} >250\,$M$_\odot$
are predicted instead to collapse to black holes without metal
ejection (HW); thus, nucleosynthetic tracers only provide a lower
limit on early star formation. The top panel shows the yields
$y_i=M_i/M_*$ (where $M_i$ is the total mass of element $i$ produced,
and $M_*$ is the initial stellar mass), assuming a He core mass of
$M_{\rm core} \approx 0.5\,M_*$ as in HW.  In zero metallicity stars
the efficiency of nuclear--powered radial pulsations (Baraffe, Heger,
\& Woosley 2001) and radiatively driven stellar winds (Kudritzki 2000)
is greatly reduced, so mass loss is likely to be small. From the
figure it is clear that iron provides an uncertain measure of VMS
formation.  The amount of silicon burned into iron in the final VMS
explosion increases with the kinetic energy of the initial collapse,
and thus the iron yield varies strongly with the core mass.  Yields of
other nuclides depend less on $M_*$. The lower panel of
Fig. \ref{yields} depicts the ratio of the mass of species $i$
produced in the VMS to the total mass of metals, divided by the
analogous ratio for solar abundances. It illustrates the non-solar
yields of VMSs. While the mass yield of oxygen is quite high, the
$\alpha$ elements Si, S, and Ca have enrichment factors relative to
solar which are far above O.
These trends are specific to very low/zero--metallicity
VMSs. The Si yield is about $\sim 10 \%$ in a $\sim 150\,\msun$
VMS which is metal-free (HW), or has a metallicity $Z=4
\times 10^{-4} Z_{\odot}$ (Portinari \etal 1998). However,
the Si yield drops by between 2 and 4 orders of magnitude for VMSs
with metallicity $Z \ge 4 \times 10^{-3} Z_{\odot}$ (Table 12 of
Portinari \etal 1998); likewise, the maximum Si yield of Type II
supernovae (SNe) from progenitor stars of any mass and metallicity is
about an order of magnitude less, $\sim 1\%$. {\it Thus, at low
metallicities Si, S, Ca abundances should be excellent tracers of
Population III VMSs.} No elements above the iron group are produced in
the HW models for the pair instability explosion.  If these models are
correct, VMSs should produce copious $\alpha$-elements with very
little associated $r$-element production.  

\begin{figurehere}
\epsscale{1.00}
\plotone{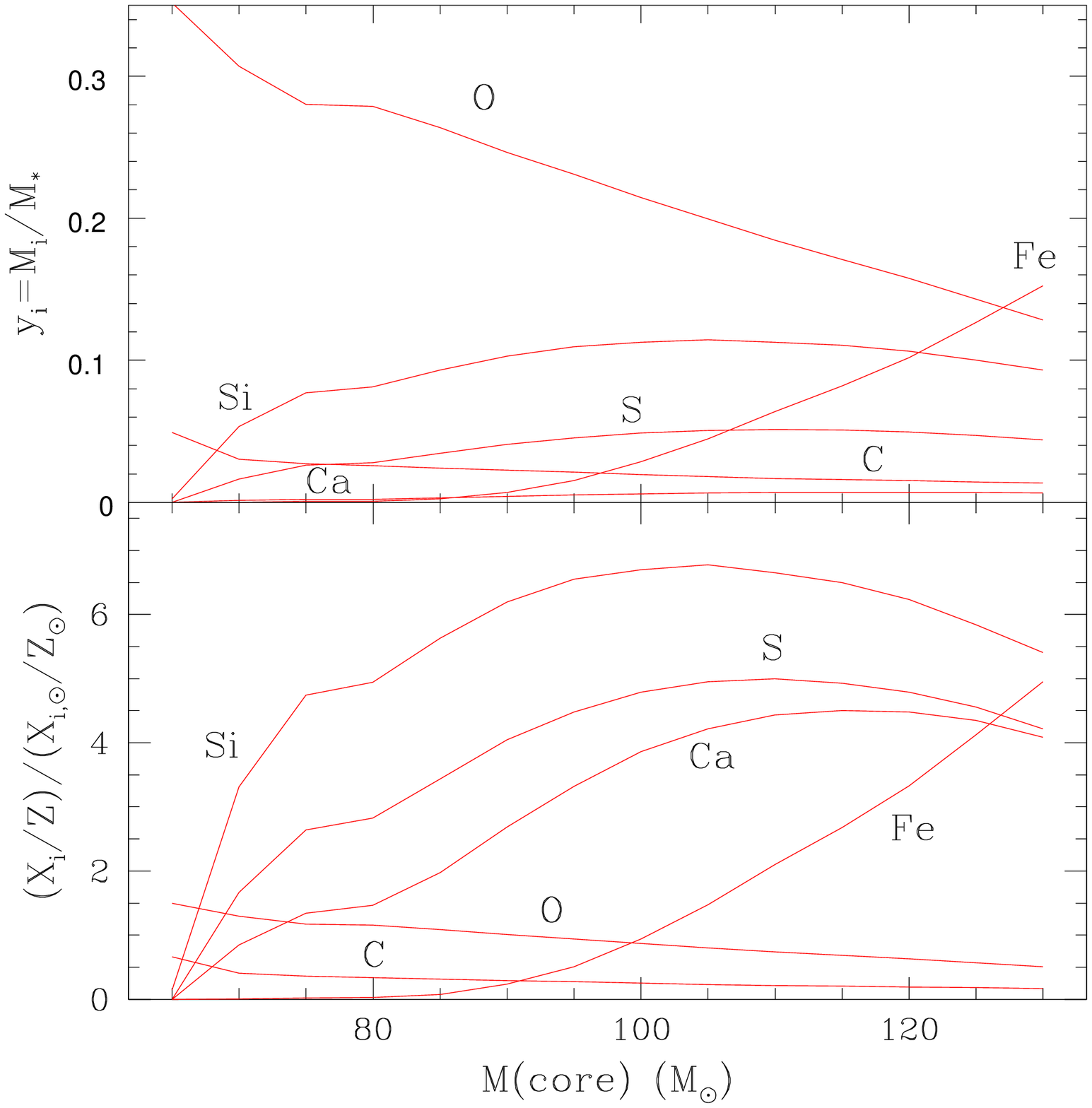}
\caption{\footnotesize VMS yields from Heger \& Woosley (2001) used in
this paper as a function of the He core mass. {\it Top panel}: the
yield $y_i=M_{i}/M_*$, where $M_{i}$ is the mass of species $i$
ejected, and $M_{*}$ is the initial stellar mass [roughly twice the
final He core mass M(core)].  {\it Bottom panel}: See text.}
\label{yields}
\vspace{+0.5cm}
\end{figurehere}

We now turn to observations. WQ analysed data from observations of heavy
$r$--process abundances in metal--poor halo stars as a function of [Fe/H] (see Fig.
\ref{rprocess}, {\it top left}). They found the following trend in the
data. At the lowest metallicities, [Fe/H]$<-3$, the Ba abundance is
roughly constant. Then, over a small range in Fe abundance, $-3.1 \le
{\rm [Fe/H]} \le 2.5$, there is a wide spread in the Ba abundance of
order 2 dex. They interpreted this sudden increase in the Ba abundance
dispersion as follows. Previous studies of radioactive $^{129}$I and
$^{182}$Hf in meteorites appear to require two distinct classes of
Type II SNe to account for $r$--process elemental abundances:
high--frequency events (H) with characteristic timescales $\sim
10^{7}$yrs that produce heavy $r$--process elements with mass numbers
$A > 130$ (e.g. Ba, Eu) and low--frequency events (L) with
characteristic timescales $\sim 10^{8}$yrs that produce light
$r$--process elements with mass numbers $A < 130$ (e.g. iodine)
(Wasserburg, Busso, \& Gallino 1996). Because of their shorter
lifetime, the H events are the first SNe to explode. As they are not
associated with significant coproduction of iron, they will produce a
wide scatter in Ba abundance over a small range in [Fe/H].  However,
the region [Fe/H]$<-3$, where the Ba abundance is very low and roughly
constant, was considered by WQ and Qian \& Wasserburg (2001b) to be an
initial or prompt inventory dominated by the contributions of very
low-metallicity massive
first-generation stars, which appear before H events and produce
very small amounts of $r$-process elements relative to their iron
production rate. In this case, the iron enrichment up to
[Fe/H]$\sim -3$ is almost exclusively due to VMSs. The finite Ba
abundance at [Fe/H]$< -3$ can be associated with very minor
coproduction of Ba in VMSs, or trace contamination by H events (see WQ and
Qian \& Wasserburg 2001b). 

\begin{figurehere}
\epsscale{1.00}
\plotone{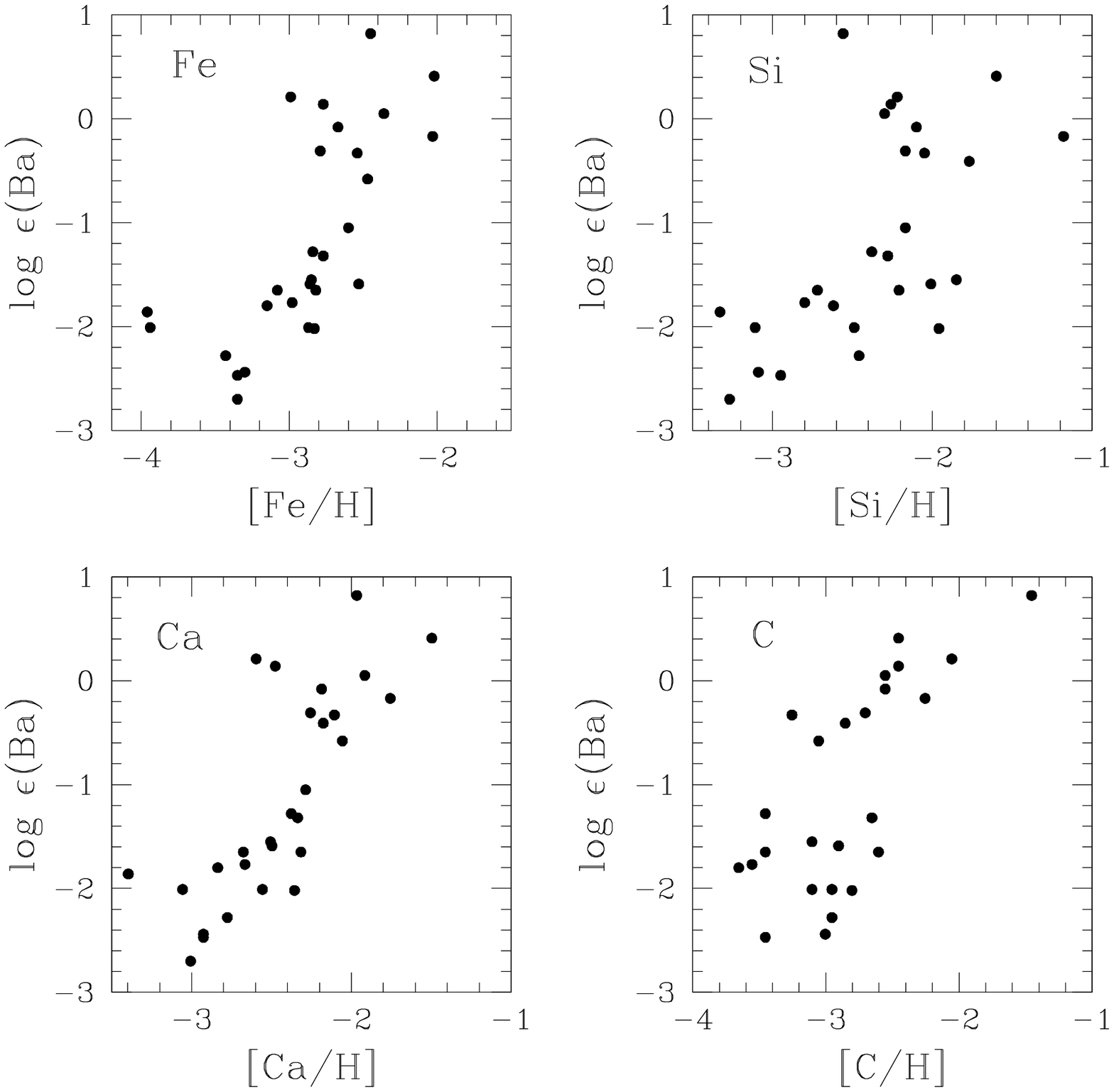}
\caption{\footnotesize Observed Ba abundances vs. other metal
abundances in metal poor halo stars, from the data of McWilliam \etal
(1995). The original trend found by Wasserburg \& Qian (2000) was in
Ba vs. [Fe/H]; the sharp increase in scatter of Ba/H at fixed
[Fe/H]$\sim -3$ also occurs in elements overproduced by VMSs, such as
Si and Ca, suggesting that the sources of Ba and Fe, Si, Ca elements
are distinct. Elements not strongly overproduced by VMSs, such as C,
do not show such a trend.}
\label{rprocess}
\vspace{+0.5cm}
\end{figurehere}

From our previous discussion we would expect the trend seen in Fe to
be most strongly exhibited in the $\alpha$ elements. In Fig.\ref{rprocess}, we plot
halo-star Ba/H values against $\alpha$ element abundances, using the
data of McWilliam \etal (1995).  Indeed, Si and Ca show the same trend
of a low Ba abundance at very low metallicity followed by a wide
scatter over a small range in metallicity. On the other hand, we would
expect these trends to be absent in elements less abundantly
produced by VMSs, which do not experience significant initial
enrichment before normal star formation takes over. This is indeed seen: carbon
shows merely a linear correlation between C and Ba production with a great deal of scatter (lower right
panel). This provides additional supporting evidence for
early VMS formation.  Most importantly, while VMS Fe yields vary
strongly as a function of mass, Si and Ca yields do not change widely
above $M_{\rm core} > 70 \msun$ ($M_{*} > 150 \msun$, from
from eqn. 1 of HW). In \S\ref{ion_photon} we will use these yields to
quantify the amount of early VMS formation. 

Although oxygen is the element produced most abundantly in VMSs,
unfortunately we cannot use it to measure VMS formation.  Oxygen
abundance estimates in low metallicity stars are uncertain, with
different trends of [O/Fe] vs. [Fe/H] claimed by different observers
(e.g., Israelian \etal 2001; Carretta \etal 2000).  Moreover, to the
best of our knowledge, there is no substantial body of simultaneous O
and $r$-process abundances in halo stars. Note that the predicted O
yields of VMSs are similar to SNII yields, and so the nucleosynthetic
signature of VMSs may be less distinct in O than in the $\alpha$
elements (see, however, Qian \& Wasserburg 2001a).

\section{Can Very Massive Stars reionize the universe?}
\label{ion_photon}

VMSs are $\sim$1-2 orders of magnitude more efficient than normal
stars with a standard IMF at producing ionizing photons, per unit
stellar mass (Bromm et al 2001c). Thus, even if the cooling and star
formation efficiencies in gas with primordial composition are low, a
much smaller fraction of the gas needs to be turned into stars in
order to reionize the universe.  We now attempt to determine if VMSs
could indeed have reionized the universe.  The onset of the large
increase in $\log\epsilon ({\rm Ba})$ occurs when the abundances of Si
and Ca are [Si/H]$_{\rm jump}$~$\approx -2.3$ and [Ca/H]$_{\rm
jump}$~$\approx -2.4$. Let us assume that the IGM was uniformly
polluted by VMSs up to this critical metallicity.  (Si abundances in
the Ly-$\alpha$ forest yield [Si/H]~$\sim -2.3$, with a scatter of a
factor $\sim 3$ [Cowie \& Songalia, 1998; Ellison et al., 2000], in
good agreement with our estimate of [Si/H]$_{\rm jump}$.)  The number
density $n_i$ of species $i$ relative to hydrogen ($n_i/n_H$) at the
critical metallicity is determined by the fraction $f_{\rm VMS}$ of
baryonic matter that was processed in VMSs.  Using $[i/H]_{\rm jump} =
{\rm log} \left( \frac{(n_{i}/n_{H})_{\rm
jump}}{(n_{i}/n_{H})_{\odot}} \right)$, we have
\begin{eqnarray}
f_{\rm
VMS}=(n_i/n_H)_{\rm jump}/(n_i/n_H)_{\rm VMS}=X_{i,\odot}\times
10^{[i/{\rm H}]_{\rm jump}}/y_i. \nonumber  
\end{eqnarray}
Here, $X_{i,\odot}$ is the mass fraction of species $i$ in the sun and
$y_i$ is the mass yield of species $i$ from a VMS. From Fig. 1, we
have $y_{\rm Si} = 0.05 - 0.1$ for all the stars with $150 M_{\odot} <
M_{*} < 250 M_{\odot}$.  We thus estimate that a fraction $f_{\rm VMS}
\sim 3- 7 \times 10^{-5}$ of all baryons must have been processed
through VMSs. Ca yields a similar range, but extending up to
$1.2\times 10^{-4}$. The result is fairly insensitive to the assumed
IMF since Ca and in particular Si yields are fairly constant for
$M_{\rm core} > 70 M_{\odot}$ ($M_{*} > 150 M_{\odot}$). An IMF
dominated by VMSs with $M_{*} < 150 M_{\odot}$ can be excluded since
these stars produce very little Fe, and the observed jump in Ba
abundance at [Fe/H]$\sim -3$ would not be seen.

In the above estimate we assume that metals produced by VMSs are
uniformly mixed throughout the IGM, and this is a fairly uncertain assumption. It is possible that early star formation was highly
biased and VMSs only polluted some small fraction of the IGM from
which the halo stars formed. Our estimate of $f_{\rm VMS}$ would then
be correspondingly reduced.  Nonetheless, metals have been detected
even in {\it underdense} gas in the IGM at $2 < z < 3$, far from star
forming regions (Schaye et al 2000), plausibly the result of early
uniform enrichment by PopIII stars.  We can use extended
Press-Schechter theory to compute the extent to which VMS debris
should be over-represented in the Galactic halo; for typical
assumptions, the enhancement factor is only $\sim 1.6$ (Madau \& Rees
2001).

How many ionizing photons does our estimated $f_{\rm VMS}$ correspond
to?  Bromm et al (2001c) have computed models of VMS spectra and find
that for $300 \msun < {\rm M_{*}} < 1000 \msun$ VMSs, the stellar
effective temperature and luminosity per solar mass are approximately
constant.  From their models, a fraction $f_{\ast,\rm BKL} \sim
(10,1,4) \times 10^{-5}$ of the baryons in the universe must have been
processed through VMSs to obtain 10 (HI,HeI,HeII) ionizing photons per
H and He nucleus in the universe respectively (see their Table 1 and
eq. 9). From this, we estimate that VMSs must have radiated $f_{\rm
photon} \sim 10 f_{\rm VMS}/f_{*,\rm BKL} \sim$(3--7, 30--70, 8--15)
photons above the (HI,HeI,HeII) ionization threshold, per H or He
nucleus. This is probably an overestimate, because the stars which
eject metals have lower masses, $150-250 \msun$, than the stars
considered by Bromm et al. (2001b).  For 100$\msun$ stars the number
of ionizing photons per solar mass is reduced by a factor $\sim 1.5-2$
for HI and HeI and $\sim 4$ for HeII (Bromm et al 2001c; Tumlinson \&
Shull 2000). The uncertainties in our estimate are almost certain to
be dominated by our assumption of uniform metal mixing, and by our
neglect of more massive stars $M_{*} > 260 \msun$ which do not eject
metals but collapse to black holes.

Could these ionizing photons escape from their host halos? Suppose
that VMSs originated in the first pregalactic objects able to cool
within a Hubble time, i.e. in halos of mass ${\rm M > M_{\rm crit} = 4
\times 10^{5} \left( \frac{1+z}{25} \right)^{-2} M_{\odot}}$, (Fuller
\& Couchman 2000; Haiman, Thoul \& Loeb 1996). Gas
within such small halos is quickly ionized by the flux from a single
VMS, and the escape fraction of ionizing photons is therefore likely to be of order unity. We obtain
the density profile of the gas $n(r)$ by assuming it is initially
isothermal at the virial temperature and demanding hydrostatic
equilibrium within an NFW dark matter profile (Makino et al 1998). The maximum
recombination rate is $\dot{N}_{\rm rec} \approx
\alpha_{B} \int n^{2}(r) 4 \pi r^{2} dr = 3 \times 10^{48} \, \left(
\frac{\sigma_{v}}{5 \, {\rm km \, s^{-1}}} \right)^{3} \left(
\frac{1+z}{25} \right)^{3/2} {\rm s^{-1}}$. This is much smaller than
the VMS HI ionizing photon production rate, $\dot{N}_{\rm ion} \approx 5 \times
10^{50} \left( \frac{M_{*}}{300 \, M_{\odot}}\right) \, {\rm
photons \, s^{-1}}$, and so the halo gas remains wholly ionized. Moreover, gas
within these small halos is easily photoevaporated: the
baryonic gravitational binding energy, $E_{b} \sim 7 \times 10^{48}
\left( \frac{M}{5 \times 10^{5} \, M_{\odot}} \right)^{5/3}
[(1+z)/20]$ erg, is considerably less than the total ionizing
radiation produced by a VMS, $E_{\rm ion}\sim 10^{54}
(M_{*}/200\,\msun)\,$erg. The entire baryonic component of the halo will be
photoevaporated on the sound-crossing timescale of the halo, $t_{\rm
evap} \sim \frac{1}{3} r_{\rm vir}/c_{s} \sim 3 \times
10^{6} \left( \frac{\sigma_{v}}{5 \, {\rm km \, s^{-1}}} \right)
\left( \frac{1+z}{25} \right)^{-3/2}$yrs, comparable to the lifetime
of the VMS, $t_{*} \sim 3 \times 10^{6}$yr. 

The amount of reionization is determined by competition between
ionization and recombination. The increased mean density of the IGM at
high redshift $n \propto (1+z)^3$ significantly increases
recombination rates. On the other hand, the gas clumping factor due to
dense non-linear objects $C \equiv \langle n_{H}^{2} \rangle/\langle
n_{H}\rangle^{2}$ declines very sharply at high redshift, and is $C
\approx 2$ at $z=20$, compared to $C \approx 30$ at $z=10$ (Haiman,
Abel \& Madau 2001).  Hence, the recombination time $t_{\rm rec} =
1/\alpha_B\,n_e C$ is in fact longer by a factor $\sim
(30/2)/(20/10)^{3} \sim 2$ at $z=20$. If (many generations of) VMSs
form over a Hubble time $t_{\rm H}$ at $z\sim 20$, then the number of
recombinations is $n_{rec} \sim t_{\rm H}/t_{\rm rec}\approx
(9,50)[(1+z)/20]^{1.5}(C/2)$ per \HII and \HeIII particle,
respectively.  Since the halo star abundances suggest that $n_{ion}
\sim (3-7,30-70,8-15)$ photons above the (HI,HeI,HeII) ionization
threshold per H or He nucleus were produced, the filling factor of
\HII and \HeIII regions due to VMSs may be $\sim n_{ion}/n_{rec} \sim
0.35-0.75$ and $\sim 0.15-0.30$ respectively.  We may have
underestimated the ionizing photon production rate and thus the
filling factors by factors of a few, since we do not count VMSs which
collapse to black holes and do not eject their metals. It is thus
plausible that the IGM was completely ionized by VMSs. Once the IGM is
reionized, a much lower comoving emissivity is required to keep it
ionized at lower redshifts than to ionize it {\it ab initio}: the high
IGM temperature strongly suppresses gas accretion onto low mass halos
$v_{c} < 30 \, {\rm km\ s^{-1}}$, so the gas clumping factor remains
small at lower redshift. This could reduce the comoving emissivity
required for an ionized IGM by a factor $C \sim 30$ at $z=10$.

Our estimate that $f_{\rm VMS} \sim 3-7 \times 10^{-5}$ is consistent
with other estimates.  Due to feedback processes, it is unlikely that
more than one VMS forms per halo (Madau \& Rees 2001).  If VMSs form
in the first pregalactic objects able to cool within a Hubble time,
with masses ${\rm M > M_{crit}= 4 \times 10^{5} \, \left(
\frac{1+z}{25} \right)^{-2} M_{\odot}}$, at a time when such halos
collapse from 2.5 $\sigma$ density peaks in the universe and contain
$\sim 1.3 \%$ of the total mass of the universe, then $f_{\rm VMS}
\sim 1.3 \times 10^{-2} \times M_{\rm VMS}/(f_{B} M_{\rm crit}) = 5
\times 10^{-5}$ (where $f_{B} \equiv \Omega_{b}/\Omega_{m}$). Normal
star formation probably starts when the gas metallicity is
sufficiently high for efficient metal-line cooling, reducing the Jeans
mass and enhancing fragmentation. Bromm \etal (2001b) find from
numerical simulations that this is likely to take place once $Z$
reaches $ 1\ {\rm to}\ 10 \times 10^{-4} Z_{\odot} \sim 2\ {\rm to}\
20 \times 10^{-6}$.  Similarly, Hellsten \& Lin (1997) find that the
critical metallicity for isobarically cooling gas at $nT\sim
10^{4}\,$K cm$^{-3}$ to cool to $T \sim 50$K is [Fe/H]$\sim -3$, the
critical metallicity identified by WQ.  Since half the main-sequence
mass of a VMS becomes metals in the HW models, our estimate of
$f_{VMS}$ corresponds roughly to $Z\sim 0.5 f_{VMS} \sim 1.5\ {\rm
to}\ 3.5 \times 10^{-5}$.  However, caution is necessary because the
cooling studies used solar metal abundances scaled with metallicity,
which are very different from the HW yields.  The difference may be
unimportant if oxygen is the dominant coolant (since O yields from HW scale with metallicity in roughly solar ratios, see Fig. 1), but a direct
comparison requires a rigorous calculation of the cooling properties
of gases with non-solar ratios of metal abundances.

Finally, due to the high effective temperature of VMSs, HeII may
reionize at high redshift in VMS reionization scenarios, only to
recombine once ordinary stars with softer spectra become predominant
(subsequent HeII reionization only took place at $z \sim 3$, as
observed by Kriss et al 2001).  In standard stellar reionization
scenarios, HeII reionization takes place only once, when quasars turn
on. The higher temperatures achievable with early HeII reionization
would have interesting consequences for the thermal history of the IGM
(Miralda-Escude \& Rees 1994; Abel \& Haehnelt 1999), and the HeII
recombination lines produced in the host halo may eventually be
observable with NGST (Oh, Haiman \& Rees 2001; Tumlinson, Giroux \&
Shull 2001).  \acknowledgements
\noindent We thank A. Heger for providing his calculations of the
metal yields from very massive stars. Support for this work was
provided by NSF grant AST-0096023 (S.P.O.), by NASA through ATP grant
NAG5--4236 (P.M.), and by NASA NAG5-10293 and Caltech Division
Contribution 8771(1082) (K.M.N. \& G.J.W.).

\end{document}